\documentclass{article}
\usepackage{icrctc07}

\newcommand{\gray}{$\gamma$-ray}
\newcommand{\grays}{$\gamma$-rays}

\newcommand{\inj}{\rm inj}
\newcommand{\pubjournal}[6] {#1, #2 {\bf #3}, #4 (#5).}

\def\apj{{\it ApJ}}

\def\aap{{\it A\&A}}

\def\physrep{{\it Phys.~Rep.}}

\def\ssr{{\it Space Sci.\ Rev.}}

\def\rpp{{\it Rep.\ Progr.\ Phys.}}
\def\ass{{\it Astroph.\ Space Sci.}}
\def\nat{{\it Nature}}

\def\mathcal{{\it }}

\def\lesssim{\mathrel{\hbox{\rlap{\hbox{\lower4pt\hbox{$\sim$}}}\hbox{$<$}}}}
\def\gtrsim{\mathrel{\hbox{\rlap{\hbox{\lower4pt\hbox{$\sim$}}}\hbox{$>$}}}}
\def\ga{\gtrsim}

\def\ie{{ i.e.,\ }}

\def\p1{{Part~I\ }}

\def\ln{{\rm ln}}

\title{Hadronic Gamma Rays from Supernova Remnants}

\shorttitle{Hadronic gamma rays from SNRs}

\abstract{ 
A gas cloud near a supernova remnant (SNR) provides a
target for $pp$-collisions leading to subsequent \gray\ emission
through $\pi^0$-decay.  The assumption of a power-law ambient spectrum
of accelerated particles with index near --2 is usually built into
models  predicting the spectra of very-high energy (VHE) \gray\
emission from SNRs.  However, if the gas cloud is located at some
distance from the SNR shock,  this assumption is not necessarily
correct.  In this case, the particles which interact with the cloud
are those leaking from the shock and  their spectrum is approximately
monoenergetic with the injection  energy gradually decreasing as the
SNR ages.  In the GLAST energy range the \gray\ spectrum resulting from particle interactions
with the gas cloud will be flatter than expected, with the cutoff
defined by the pion momentum distribution in the  laboratory frame.
We evaluate the flux of particles escaping from a SNR shock and apply
the results to the VHE diffuse emission detected by the HESS at the
Galactic centre.  
}

\authors{I. V. Moskalenko$^{1,2}$, T. A. Porter$^3$, M. A. Malkov$^4$, P. H. Diamond$^4$}
\shortauthors{Moskalenko et al.}
\afiliations{
$^1$Hansen Experimental Physics Laboratory, Stanford University, Stanford, CA 94305, U.S.A.\\
$^2$Kavli Institute for Particle Astrophysics and Cosmology, Stanford University, CA 94309, U.S.A.\\
$^3$Santa Cruz Institute for Particle Physics, University of California, Santa Cruz, CA 95064, U.S.A.\\
$^4$University of California at San Diego, La Jolla, CA 92093, U.S.A.}
\email{imos@stanford.edu}

\begin{document}
\maketitle


\section{Introduction}

SNRs are believed to be the primary sources of cosmic rays (CR) in the Galaxy.
Observations of X-ray \cite{Koyama1995} and \gray\ emission 
\cite{Aharonian2005b,Aharonian2006}
from SNR shocks reveal the presence of energetic particles,
thus testifying to efficient acceleration processes.
Acceleration of particles in collisionless shocks is a matter of
intensive research in conjunction with the problem of CR origin 
\cite{Drury1983,Blandford1987,Jones1991}.
Current models include nonlinear effects
(e.g., \cite{Berezhko2006}) and treat particle
acceleration using hydrodynamic codes 
(e.g., \cite{Ellison2005}).
The predicted spectrum of accelerated
particles has a power-law form in rigidity with index which may
slightly vary around --2.

The VHE \gray\ emission from shell-type SNRs has been modelled using 
leptonic (inverse Compton -- IC) and hadronic ($\pi^0$-decay) scenarios.
The leptonic scenario
fits the broad-band spectrum of a SNR assuming a pool of accelerated
electrons IC scattering off the interstellar radiation field
producing VHE \grays\ while the magnetic field and electron spectrum cut-off 
are tuned to fit the radio and X-ray data 
(e.g., \cite{Lazendic2004,Porter2006}). 
The hadronic model fits the VHE \gray\
spectrum assuming a beam of accelerated protons hits a target, such as a 
nearby molecular cloud \cite{Aharonian2002,Malkov2005}. 
The latter, if definitively
proven, would be the first experimental evidence of proton acceleration
in SNRs.

The assumption of a power-law ambient spectrum of accelerated
particles with index near --2 is usually built into the models 
predicting the spectra of VHE \gray{} emission from SNRs.
However, if a molecular cloud is located
at some distance from the shock, the particles which
interact with the cloud are those leaking from the shock and their spectrum
is different \cite{Gabici}. 
In a toy model, the shock accelerates
particles until the highest possible energy is reached. 
At this point
the shock cannot confine the particles any longer and they escape
into the interstellar medium (ISM). 
The energy spectrum of these particles will be
monoenergetic with the injection energy gradually decreasing as the SNR ages.
The \gray\ spectrum resulting from the particles interacting
with a gas cloud will be essentially flatter than expected. 
The diffuse emission detected at the Galactic centre by the HESS 
\cite{Aharonian2006a} may be of this sort.

\section{Particle acceleration in SNR shock\label{model}}

We use the steady-state diffusion-convection equation
\begin{equation}
u\frac{\partial f}{\partial x}+\kappa(p)\frac{\partial^{2}f}{\partial x^{2}}=
\frac{1}{3}\frac{du}{dx}p\frac{\partial f}{\partial p},
\label{c:d}
\end{equation}
where $f(x,p)$ is the isotropic (in the local fluid frame) part
of the particle distribution, and $p$ is the particle momentum in $mc$ units. 
We use a planar geometry and assume that the gaseous discontinuity
(sub-shock) is located at $x=0$ while the shock propagates in the positive
$x$-direction. 
The flow velocity in the shock frame can be
represented as $V(x)=-u(x)$ where the (positive) flow speed $u(x)$
jumps from $u_{2}\equiv u(0-)$ downstream to $u_{0}\equiv u(0+)>u_{2}$
across the sub-shock and then gradually 
increases up to $u_{1}\equiv u(+\infty)\geq u_{0}$.
The particle density is assumed to vanish far upstream
($f\rightarrow0,\, x\rightarrow\infty$), while the only bounded solution
downstream is $f(x,p)=f_{0}(p)\equiv f(0,p)$. 
We assume
Bohm diffusion $\kappa(p)=Kp^{2}/\sqrt{1+p^{2}}$.
The constant $K$ depends on the $\delta B/B$ level of the 
magnetohydrodynamic turbulence that
scatters particles in pitch angle.
However, it can be rescaled
out of eq.~(\ref{c:d}) since we are not interested in the shock structure here. 

To include the back-reaction of accelerated particles on the plasma
flow the following equations are used in a quasi-stationary acceleration
regime: (i) the conservation of the momentum flux in the
smooth part of the shock transition (CR-precursor, $x>0$) 
$P_{\textrm{c}}+\rho u^{2}=\rho_{1}u_{1}^{2}$,
where $P_{\textrm{c}}$ is the CR pressure (CRs escape after crossing
$p=p_{\max}$), (ii) the continuity equation $\rho u=\rho_{1}u_{1}$,
(iii) the Rankine-Hugoniot relations for the sub-shock strength 
$r_{s}\equiv u_{0}/u_{2}=\left(\gamma+1\right)/\left(\gamma-1+2R^{\gamma+1}M^{-2}\right)$
where $M$ is the Mach number at $x=\infty$, the precursor compression
$R\equiv u_{1}/u_{0}$, and $\gamma$ is the adiabatic index of the
plasma. 

These equations self-consistently describe the particle
spectrum and flow structure. 
An efficient solution method is to 
reduce this system to one integral equation \cite{MDru01}. 
A key dependent
variable is an integral transform of the flow profile $u(x)$ with
a kernel suggested by an \emph{exact} (in the limit $M\to\infty$,
$p_{\max}\to\infty$) asymptotic solution of the system which has the
following form 
\begin{equation}
f(x,p)=f_{0}(p)\exp\left[-\frac{q}{3\kappa}\Psi\right],
\label{eq:sol}
\end{equation}
where the flow potential $\Psi=\int_{0}^{x}u(x')dx'$ and the spectral
index at the sub-shock and downstream $q(p)=-d\ln f_{0}/d\ln p$.
The integral transform generates the {}``spectral function'' of
the flow velocity $U\left(p\right)$ instead of the flow velocity
$u\left(x\right)$ as follows 
\begin{equation}
U(p)=\frac{1}{u_{1}}\int_{0-}^{\infty}\exp\left[-\frac{q(p)}{3\kappa(p)}\Psi\right]du(\Psi),
\label{U}
\end{equation}
and is related to $q(p)$ as
$q(p)=d\ln U/d\ln p+3/[r_{s}RU(p)]+3$.
The equation for $U\left(p\right)$ is
\begin{eqnarray}
U(t) &\!\!\!\!\!=\!\!\!\!\!& \frac{r_{s}-1}{Rr_{s}}\!+\!
\frac{\nu}{Kp_{0}}\int_{t_{0}}^{t_{1}}\!\!dt'\!\!\left[\frac{1}{\kappa(t')}\!
+\!\frac{q(t')}{\kappa(t)q(t)}\right]^{-1}\nonumber \\
 &\!\!\!\!\!\times\!\!\!\!\!&\frac{U(t_{0})}{U(t')}\exp\left[-\frac{3}{Rr_{s}}
\int_{t_{0}}^{t'}\frac{dt''}{U(t'')}\right],\label{int:eq}
\end{eqnarray}
where $t=\ln p$, $t_{0,1}=\ln p_{0,1}$. The injection parameter
$\nu=(4\pi/3)(mc^{2}/\rho_{1}u_{1}^{2})p_{0}^{4}f_{0}(p_{0})$
is related to $R$:
\begin{eqnarray}
\nu &\!\!\!\!\!=\!\!\!\!\!& Kp_{0}\left(1-R^{-1}\right)\nonumber\\
 &\!\!\!\!\!\!\!\times\!\!\!\!\!\!\!& 
\left\{ \!\int_{t_{0}}^{t_{1}}\!\!\!\!\!dt\,\kappa(t)\frac{U(t_{0})}{U(t)}\exp\!
{\left[\!
-\!\frac{3}{Rr_{s}}\!\int_{t_{0}}^{t}\!\!\frac{dt'}{U(t')}\!
\right]
}\right\}^{\!\!-1}
\label{nu}
\end{eqnarray}

\begin{figure}[t]
\centerline{
\includegraphics[%
  clip,width=2.5in]
{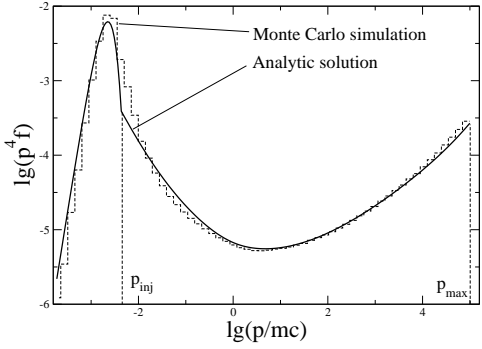}}
\caption{Comparison of our analytic solution with MC simulation \cite{EllisBerBar00}.
Below $p_{\inj}$, it is a thermal Maxwellian; 
MC simulation parameters: $M$$=$$128$, $p_{\max}$$=$$10^{5}$, $\nu$$\simeq$$0.4$.
\label{fig:comp}}
\end{figure}

\begin{figure}[t]
\centerline{
\includegraphics[%
  clip,width=2.5in]
{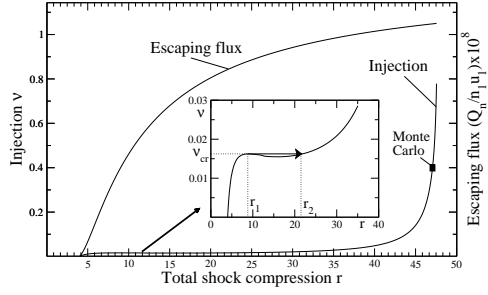}}
\caption{Bifurcation diagram showing injection rate $\nu$ (which is actually
a control parameter) versus $r$. 
The insert shows the region of the ``phase transition.''
The upper curve shows the particle
escaping flux, normalised to the plasma flux far upstream and multiplied
by the factor $10^{8}$, for clarity.\label{cap:Bifurcation-diagram}}
\end{figure}

Our main goal is to calculate the flux of escaping particles.
Since they leave the accelerator through the $p=p_{\max}$ boundary
and the solution normalisation is set by the particle distribution
at $p=p_{\inj}$ (injection rate), the flux will depend on the entire
solution between $p_{\inj}$ and $p_{\max}$. 
For typical SNR conditions,
this interval spans seven orders of magnitude or more and even small
errors in the slope along the spectrum will result in a failure of
the flux determination. 
To avoid this, we first compare our analytic
solution with a numerical one \cite{EllisBerBar00}. 
The result of
the comparison is illustrated in Fig.~\ref{fig:comp}. 
To
compare we substitute the injection rate (\ie the height of
the spectrum at $p=p_{\inj}$) indicated in the Monte Carlo (MC) simulations.
To calculate the escaping flux we use the time dependent
version of eq.~(\ref{c:d}) using $g\equiv p^{3}f$ in place of $f$
\begin{equation}
-\frac{\partial g}{\partial t}+\frac{\partial}{\partial x}\left(ug+\kappa(p)\frac{\partial g}{\partial x}\right)=\frac{1}{3}\frac{du}{dx}p\frac{\partial g}{\partial p},\label{eq:dctd}
\end{equation}
and calculate the temporal variation of the total number of particles
in the region $x>0$ assuming $p_{\max}\simeq const$ (see \cite{MD06} for a 
discussion of this assumption, along with 
our assumption about monoenergetic escape and for further references). 
Note that $p_{\max}(t)$ is adopted in \cite{Ptuskin,Gabici}
\[
N_{Up}=4\pi\int_{0}^{\infty}dx\int_{p_{\inj}}^{p_{\max}}g\left(p,x\right)dp/p\]
Therefore, from eq.~(\ref{eq:dctd}) we obtain
\begin{eqnarray}
\frac{\partial N_{Up}}{\partial t}
&\!\!\!\!=\!\!\!\!&\frac{4\pi}{3}\!\!\int_{0-}^{\infty}\!\!dx\frac{du}{dx}\left[g\left(p_{\inj},x\right)\!
-\!g\left(p_{\max},x\right)\right]\nonumber\\
&\!\!\!\!-\!\!\!\!&4\pi u_{2}\int_{p_{\inj}}^{p_{\max}}\!dp\,\frac{g\left(p,0-\right)}{p}.
\label{eq:bal}
\end{eqnarray}
In this particle balance equation the term containing $g\left(p_{\inj}\right)$
is simply due to particle injection at $p=p_{\inj}$, while the last
term is the particle convective flux downstream. 
The remaining term
is the flux of escaping particles 
\begin{eqnarray}
Q_{n}&=&\frac{4\pi}{3}\int_{0-}^{\infty}du\left(x\right)g\left(p_{\max},x\right)\nonumber\\
&\simeq&\frac{4\pi}{3}f_{0}\left(p_{\max}\right)p_{\max}^{3}u_{1}U\left(p_{max}\right),
\label{eq:Qn}
\end{eqnarray}
where we have used definition of the spectral function $U\left(p\right)$
eq.~(\ref{U}), and the solution, eq.~(\ref{eq:sol}). 
Using the plasma number density $n_1=\rho_1/m_p $
one can obtain the
normalised flux as
\[
\frac{Q_{n}}{n_{1}u_{1}}=\nu\frac{u_{1}^{2}}{c^{2}}\frac{p_{\max}^{3}f_{0}\left(p_{\max}\right)U\left(p_{\max}\right)}{p_{\inj}^{4}f_{0}\left(p_{\inj}\right)}.\]
This quantity is plotted in Fig.~\ref{cap:Bifurcation-diagram}
as a function of compression $r$. 
Note that the injection rate obtained
from MC simulations corresponds to nearly the maximum escaping flux
possible, where the function $r\left(\nu\right)$ saturates. 
This
saturation is related to the sub-shock smearing, $R^{\gamma+1}$$\to$$M^{2}$.
We believe that the MC method \emph{overestimates} the injection due to
the lack of the feedback from particle driven turbulence.
It was argued \cite{MDru01} that self-regulation of acceleration,
which includes but is not limited to the particle trapping by self-generated
waves, leads to a significant reduction of the injection rate which
is maintained at nearly the critical level where the shock compression
$r\left(\nu\right)$ rises sharply, Fig.~\ref{cap:Bifurcation-diagram}.
For the calculations shown in Fig.~\ref{cap:Bifurcation-diagram}, this
approach indeed reduces the injection by an order of magnitude compared to 
MC results. 

Detailed calculations of the escaping flux will be presented
elsewhere. 
Here we note that the normalised flux $Q_{n}/n_{1}u_{1}$
does not depend strongly on $M$ saturating (when $M$$\ga$$100$) at 
$Q_{n}/n_{1}u_{1}\approx10^{-5}/p_{\max}$
for $u_{1}$$=$$0.005c$. 
This scaling is the result of energy
requirement, since $Q_{e} = cp_{\max}Q_{n}$ and $Q_{e}$ is the
escaping energy flux which is constrained by the available mechanical
energy flux and the equipartition condition. 
The latter means
that for the injection rate close to the critical, about half of the
shock energy goes into the CR.

\begin{figure}
\centerline{
\includegraphics[width=2.6in]{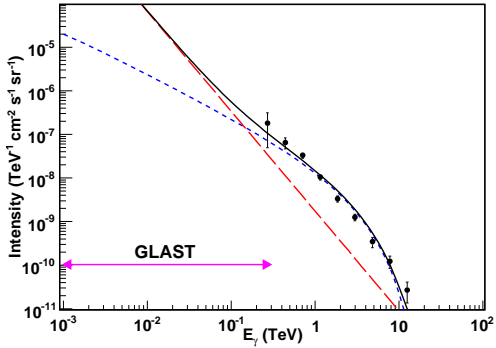}
}
\caption{The spectrum of \grays\ from outer (monoenergetic protons
of 25 TeV, blue-dots) and inner clouds (power-law with index $-2.29$,
red-dashes); normalisations are arbitrary.
Solid line shows the total spectrum. Data: HESS \cite{Aharonian2006a}.
\label{cap:gamma-ray-flux}}
\end{figure}

\section{Calculations}
Diffuse emission has been detected at the Galactic centre by the HESS 
\cite{Aharonian2006a}. 
Its intensity correlates
with the gas density as traced by the CS emission up to the $>$200 pc from
the central source, while the SNR shell is about half this size, $\sim$100 pc,
assuming a SNR age $\tau_{\rm SNR}$$\sim$10 kyr and shock 
speed $10^4$ km s$^{-1}$. 
Clearly, some part of this emission $\ga$100 pc may be produced by particles
which left the SNR shock $\tau$$<$1.5 kyr, if the diffusion coefficient in the 
ISM $Dxx(10\ {\rm TeV})$$\ga$1 kpc$^2$ Myr$^{-1}$, i.e.\ relatively
recently $\tau\ll\tau_{\rm SNR}$.  

The HESS data are obtained for the combined emission from all gas
clouds in the region.  However, the model predicts different VHE
\gray\ spectra from clouds at different distances from the SNR.  The
most distant clouds should exhibit a flatter spectrum with   index
close to --1; closer to the SNR shell the spectrum may  steepen and
become a regular power-law with index about --2 for  clouds located
at or inside the shell.  To evaluate the feasibility of such scenario,
we calculate the spectrum of VHE \grays\ assuming two components: one
for the clouds outside of the shell  (a monoenergetic ambient spectrum
of protons) and another for  the clouds inside the shell with a
power-law index similar to the central source --2.29.  The spectrum of
\grays\ can be evaluated using the scaling approximation
\cite{StephensBadhwar81} as described in \cite{MS98}.

To illustrate the idea, Fig.~\ref{cap:gamma-ray-flux} shows the
calculated two-component spectrum from the Galactic centre together
with the HESS data.  Reasonable agreement with the data can be
obtained for a proton injection energy $\sim$25 TeV, consistent with
the SNR age of $\sim$10 kyr. In the GLAST energy range the spectrum from the outer clouds
differs significantly from what is expected for the usual hadronic
scenario: it has slope close to --1.  Gabici and Aharonian
\cite{Gabici} came to a similar conclusion.

{\bf Acknowledgements.}
We thank Pasquale Blasi and Don Ellison for many useful discussions.
I.\ V.\ M.\ thanks NASA APRA grant for partial support.
T.\ A.\ P.\ is supported in part by the US Department of Energy.
M.\ A.\ M\ and P.\ H.\ D.\ were supported by NASA under grant ATP03-0059-0034 and by
the U.S. DOE under Grant No. FG03-88ER53275.


\end{document}